\documentclass{PoS}

\title{Are bottom PDFs needed at the LHC?}

\ShortTitle{b PDFs at the LHC}

\author{\speaker{Maria Ubiali}\thanks{Work in collaboration with Fabio Maltoni, Giovanni Ridolfi 
and Matthew Lim.}\\
      DAMTP, Centre for Mathematical Studies, Wilberforce Rd, CB3 0WA\\
       \& Department of Physics, Cavendish Laboratory, J.J. Thomson Avenue, CB3 0HE\\ 
       University of Cambridge, Cambridge, UK\\
        E-mail: \email{ubiali@hep.phy.cam.ac.uk}}

\abstract{Processes involving bottom quarks play a crucial role in the LHC phenomenology, from
flavour physics to Higgs characterisation and as a window to new physics, appearing both as signals 
and irreducible background in BSM searches.
These processes can be described in QCD either 
in a 4-flavor or 5-flavor scheme. In the former, $b$ quarks appear only in the final state and are 
considered massive. In 5-flavor schemes, calculations include $b$ quarks in the initial state. 
Possibly large logarithms originating from the collinear splitting of gluons into bottom pairs 
are resummed into the $b$ parton distribution function (PDF). 
In this contribution, I describe a simple method to assess the size of the logarithms in processes initiated 
by bottom quarks and show how a substantial and justified agreement between calculations
in the two schemes can be achieved. 
As a consequence both calculations can be used in different context. To conclude,
an overview of the current studies aiming to generalise the current appraisal is given 
and some preliminary results are discussed.}

\FullConference{XXII. International Workshop on Deep-Inelastic Scattering and Related Subjects,\\
		28 April - 2 May 2014\\
		Warsaw, Poland}

\begin{document}

Processes that feature gluon splitting into $b\bar b$ pairs in the 
initial state are of great phenomenological interest
at the LHC. Many SM analyses, such as the precise measurement of the 
electroweak coupling in single top production
or the Higgs boson characterisation via the study of its coupling
to bottom quarks, as well as  
searches for new physics, consider final state signatures which
involve bottom quarks. The need for accurate predictions is strongly motivated.\\
There are two different ways to compute theoretical predictions
for this class of processes, often referred to as schemes: 
the 4- and the 5-flavor schemes. 
In the former (4FS), bottom quarks 
are treated as massive particles which appear only in the final state.
The bottom quark has a non-zero transverse momentum already at the
leading order
and the full kinematics of the bottom quark is accurately described in the
next-to-leading order computations. 
Moreover the implementation of the calculation 
in parton shower codes is straightforward,
 as there is no arbitrariness in the description of massive
effects. However, order by order in the matrix element, 
logarithms of the ratio $\mathcal{Q}^2/m_b^2$, $\mathcal{Q}^2$
being of the order of the hard scale of the process, appear
as a result of the collinear splitting of
gluons into bottom pairs. 
These possibly large logarithms may spoil the convergence of the perturbative
series. 
A way out is given by the so-called 5-flavor scheme (5FS), in which
bottom PDFs are introduced. The mass of the bottom quarks is
considered as a small parameter and the bottom quarks 
contribute to the proton wavefunction and to the
running of the strong coupling constant. 
This way, the DGLAP evolution of the 
bottom PDFs from threshold to the hard scale of the process
automatically resums the initial state collinear logarithms
to all orders in perturbative QCD. 
Calculations in the 5FS always start with at least a power of $\alpha_s$ 
less than the corresponding 4FS calculations
and the bottom quarks have zero transverse momentum at leading order.
Some improved versions of the 5-flavor scheme have been proposed in which
the effects of the mass of the bottom quarks are included by replacing 
higher order contributions with no $b$ quarks in the initial state by
corresponding contributions with $m_b\neq 0$~\cite{Aivazis:1993pi}.
In principle the two schemes can be combined and consistently matched
via an approach based on Ref.~\cite{Cacciari:1998it}, in which the massive 
calculation in the 4FS is supplemented by the resummation of the initial state
collinear logarithms and double
counting is properly taken into account. However matched resummed 
calculations are available only for 
a limited number of processes at colliders and they are more
difficult to be
implemented in the description of exclusive observables. 
Therefore, in practice, total cross section
predictions in the two schemes
are often combined via some pragmatical matching prescription,
not based on a thorough field-theoretic analysis, such as the one proposed 
in~\cite{Harlander:2011aa}, currently adopted by the LHC Higgs working group. 

Clearly, if all perturbative orders were included, the 4-flavor
and the improved 5-flavor schemes would yield 
identical results. However the terms in the perturbative expansion
are organised differently and
at any finite order in perturbative QCD results are different.
In the past the discrepancy between the predictions formulated
in the two schemes appeared to be extremely large, a glaring example
being the factor of 10 difference in the case of the 
leading order $b$--initiated Higgs 
production.
Several studies have been performed to investigate 
its origin, e.g.~\cite{Dittmaier:2003ej,Maltoni:2003pn}. 
In all cases of study, increasing the perturbative order of 
the predictions did obviously help in reducing the discrepancy, 
but this was not enough to reconcile 
the results obtained in the two schemes unless they
were compared by using a factorisation scale smaller 
than the typical hard scale of the process. 
Also, contrary to one's na\"ive expectations that the 
collinear logarithms would have more space to develop 
at large energies, the discrepancy
between predictions obtained  the two schemes 
was found to be larger at the Tevatron than at the LHC.

Despite the effort invested in investigating this issue, we 
still were not able to answer a number of crucial questions:
according to which criterion can one establish which is the 
best way for describing bottom-initiated processes at colliders? 
What is the typical size of the effects of the resummation of
initial-state collinear logs of the type $\log\frac{{\cal Q}^2}{m_b^2}$
with respect to an approximation where only logs at a finite order
in perturbation theory are kept? Also, what is the typical size of 
logarithms themselves in phenomenologically relevant processes at the LHC? 
Finally, what justifies the use of a smaller factorisation
scale when comparing the 5FS predictions to the 4FS ones?

\begin{figure}[ht]
\begin{center}
\includegraphics[width=0.5\textwidth]{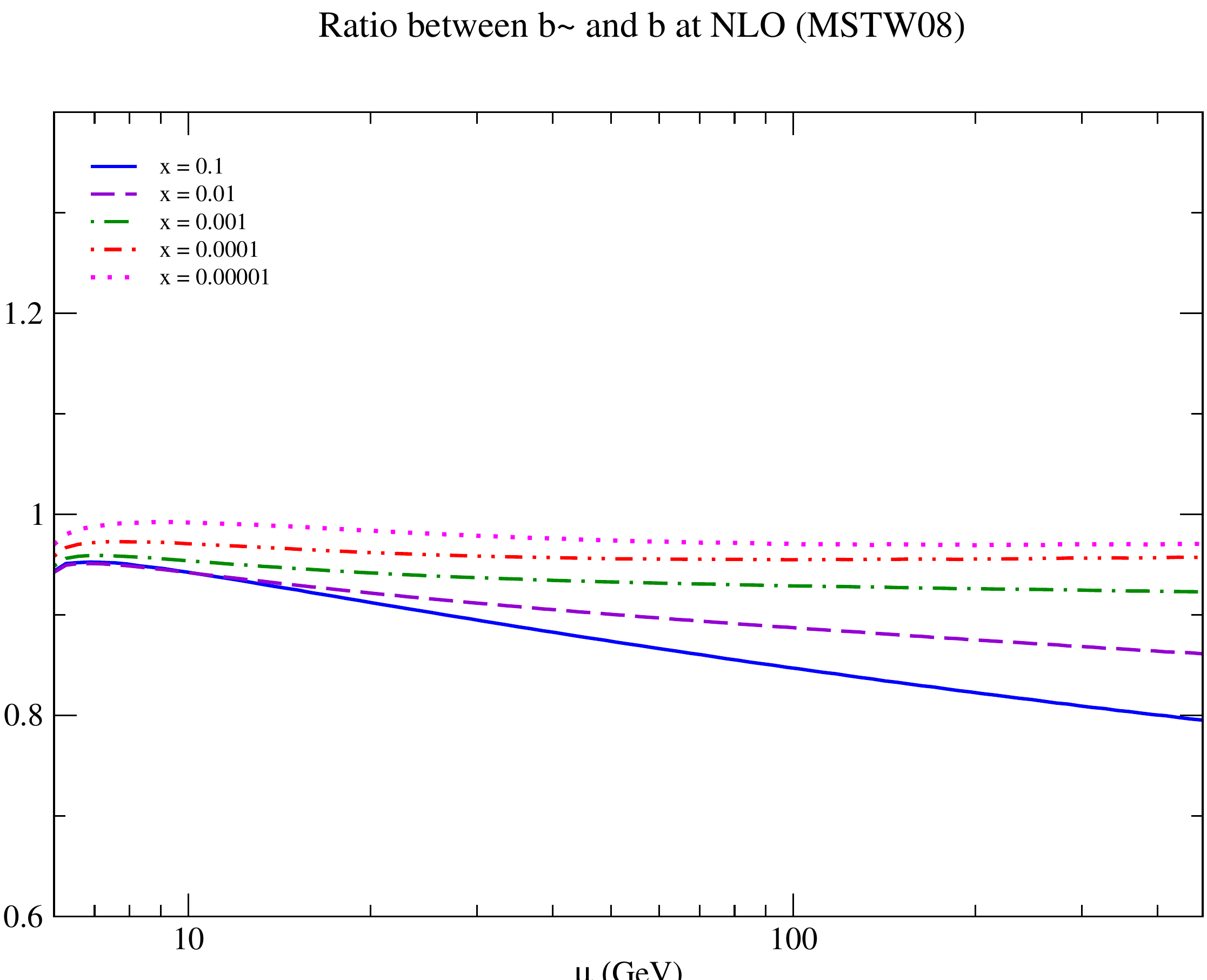}\\
\end{center}
\caption{Ratio $\tilde{b}/b$ for several values of $x$ as a function of the scale $\mu$. 
The 4F-FFNS and GM-VFNS are associated to the $\tilde{b}$ and $b$ PDF computations respectively at 
NLO order for the {\tt MSTW2008} parton set. 
\label{btilde}}
 \end{figure}

In Reference~\cite{Maltoni:2012pa} a reappraisal was 
formulated, which answered to the above questions by 
keeping into account two simple facts, one of dynamical and one of
kinematical origin. 
The first one concerns the evolution of the bottom
PDFs. In Fig.~\ref{btilde}, taken from~\cite{Maltoni:2012pa}, 
the ratio of $\tilde{b}^{(2)}$, i.e. of the approximated $b$
distribution that one obtains when truncating the perturbative
expansion of the bottom PDF evolution at $\mathcal{O}(\alpha_s^2)$,
and the full $b$ PDF obtained by solving the DGLAP evolution at
next-to-leading order, is displayed. One can observe that
the effects of the resummation of the 
$\log\frac{\mu^2_f}{m_b^2}$'s 
is quite small and relevant mainly at large Bjorken $x$.
So, in general, keeping only the explicit logs appearing at NLO is 
a good approximation and it stops being good only when the mass
of the produced particle is very large as compared to the centre of mass energy.  
This observation accounts for previously noticed behaviours, such as the more 
sizable differences between predictions
in the two schemes for single top and $b b \to H$ at the Tevatron
than at the LHC~\cite{Campbell:2009ss,Campbell:2004pu}.
Recently a similar study was performed to assess the size of 
the logarithms resummed in top PDFs at future 100 TeV colliders with very similar findings~\cite{Dawson:2014pea}.
To conclude, unless the typical Bjorken $x$ probed by the process is large, the
effects of initial-state collinear logs is always
modest, and, even though total cross sections computed in 5-flavor
schemes may indeed display a smaller uncertainty, such logarithms do
not spoil the convergence of perturbation theory in 4-flavor scheme
calculations. 

Furthermore, we showed that the effective scale
${\cal Q}$ which enters the initial-state collinear logarithms, while
proportional to hardest scale(s) in the process, turns out to be
modified by universal phase space factors. The latter are valid at all orders and
are independent of the details of the splitting. 
Considering the $m_b^2\to 0$ limit of the lowest-order 4FS cross-sections, which present a collinear
singularity due to the gluon splitting into $b\bar b$ pair in the initial state,
we have shown that the  
logarithmically-enhanced contributions to the cross-sections are proportional to
\begin{equation}
L=\log\frac{{\cal Q}^2(z)}{m_b^2},
\label{eq1:general}
\end{equation}
where ${\cal Q}^2(z)$ is a universal factor, a function of the mass of the 
produced particles $M^2$ and the virtuality of the exchanged bosons in the $t-$channel
$Q^2$ as well as their ratio $z$ with the partonic centre-of-mass energy $\hat{s}$:
\begin{equation}
{\cal Q}^2(z)=(M^2+Q^2)\frac{(1-z)^2}{z}\frac{1}{1-\frac{zQ^2}{M^2+Q^2}}
\qquad{\rm with}\qquad 
z=\frac{M^2+Q^2}{\hat{s}+Q^2}, 
\label{eq1:mu2}
\end{equation}
The collinear logarithm reduces to $L_{\rm DIS} = \log\left[\frac{Q^2}{m_b^2}\frac{1-z}{z}\right]$ 
in the case of DIS $b\bar b $ production ($M^2\to 0$)
and to $L_{\rm DY} = \log\left[\frac{M^2}{m_b^2}\frac{(1-z)^2}{z}\right]$ 
in the case of Drell-Yan production ($Q^2\to 0$).
The explicit scale ${\cal Q}^2$ is different from the one the we would naively expect
$M^2+Q^2$, i.e., of the hard scale to the collinear regulator, to develop in the integrated 
cross section.\\ 
Interestingly, the universal phase space factor 
tends to reduce the size of the logarithms for processes taking place 
at hadron colliders, while it enhances them 
in the case of DIS. In particular the suppression observed at the LHC 
is stronger than at the Tevatron and it gets stronger the heavier is the mass 
of the produced particle. By weighting the logarithm on the events according to
 a simple analytic formula, one can derive
the factorisation scale at which to perform comparisons between calculations
in the two schemes. For the processes that have been analysed,
single top production and vector boson associate production, 
the scale turned out to be similar to the scales used in 
previous phenomenological analyses, about $m_t/4$ for single top
production and $m_W/3$ for $Wb$ associated production. 
However its origin is now properly physically-motivated. 
As a result, a consistent and quantitative
explanation is provided of the many examples where a substantial
agreement between total cross sections obtained at NLO (and beyond) in
the two schemes can be found within the expected uncertainties.
A recent study of the heavy Charged Higgs production
cross section in the two schemes, which has been preliminary 
presented in~\cite{Heinemeyer:2013tqa} and which is soon
going to be published, reinforces the findings of~\cite{Maltoni:2012pa}.
The choice of the factorisation scale in the 5-flavour scheme 
calculation driven by Eq.~(\ref{eq1:mu2}) significantly improves 
the agreement between predictions in the two schemes, 
and leads to a reliable NLO QCD prediction for heavy charged Higgs boson production 
which can be used in the searches at the second run of the LHC.

To answer the question that opens this contribution: do we need $b$-PDFs at the LHC?
The main outcome of our study is that 4- and 5-flavor 
schemes provide complementary information. It is therefore strongly motivated having
calculations at higher orders available in both schemes for any given
process. (Improved) 5-flavor schemes, for example, can typically
provide quite accurate predictions for total rates and being simpler,
in some cases allow the calculations to be performed at NNLO, such as
those already available for $bb\to H,Z$. On the other hand,
being often the effects of resummation very mild, 4-flavor
calculations can be also employed. They can be useful to achieve
accurate fully exclusive predictions, such as those obtained from
Monte Carlo programs at NLO accuracy. Promoting a 4-flavor calculation
at NLO to an event generator is nowadays a fully automatic procedure
and kinematic effects due to the $b$ quark mass can be taken into
account from the start leading for example to a more accurate
description of the kinematics of the spectator $b$ quarks in all phase
space.

\begin{figure}[ht]
\begin{center}
\includegraphics[width=0.4\textwidth]{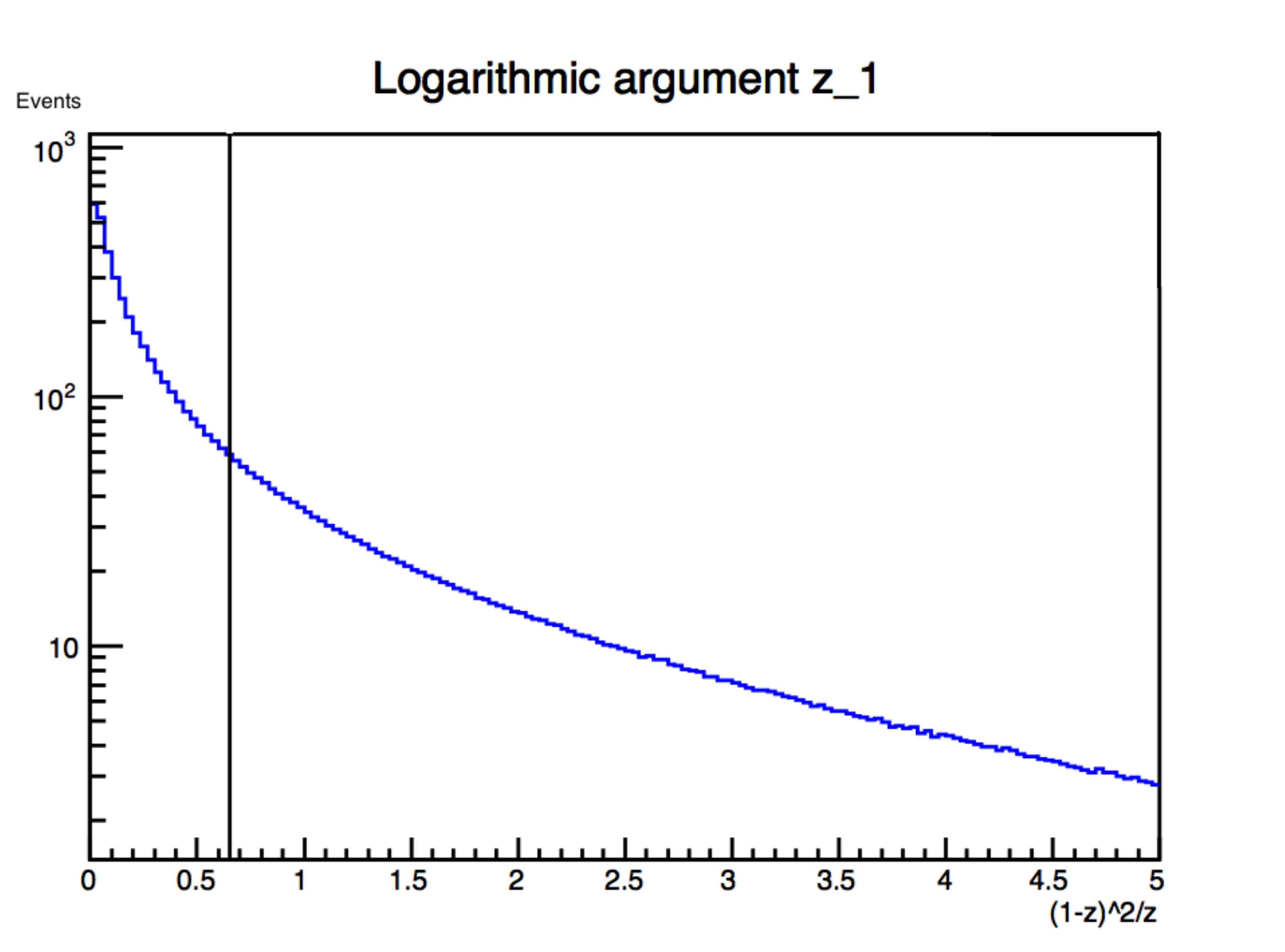}
\includegraphics[width=0.4\textwidth]{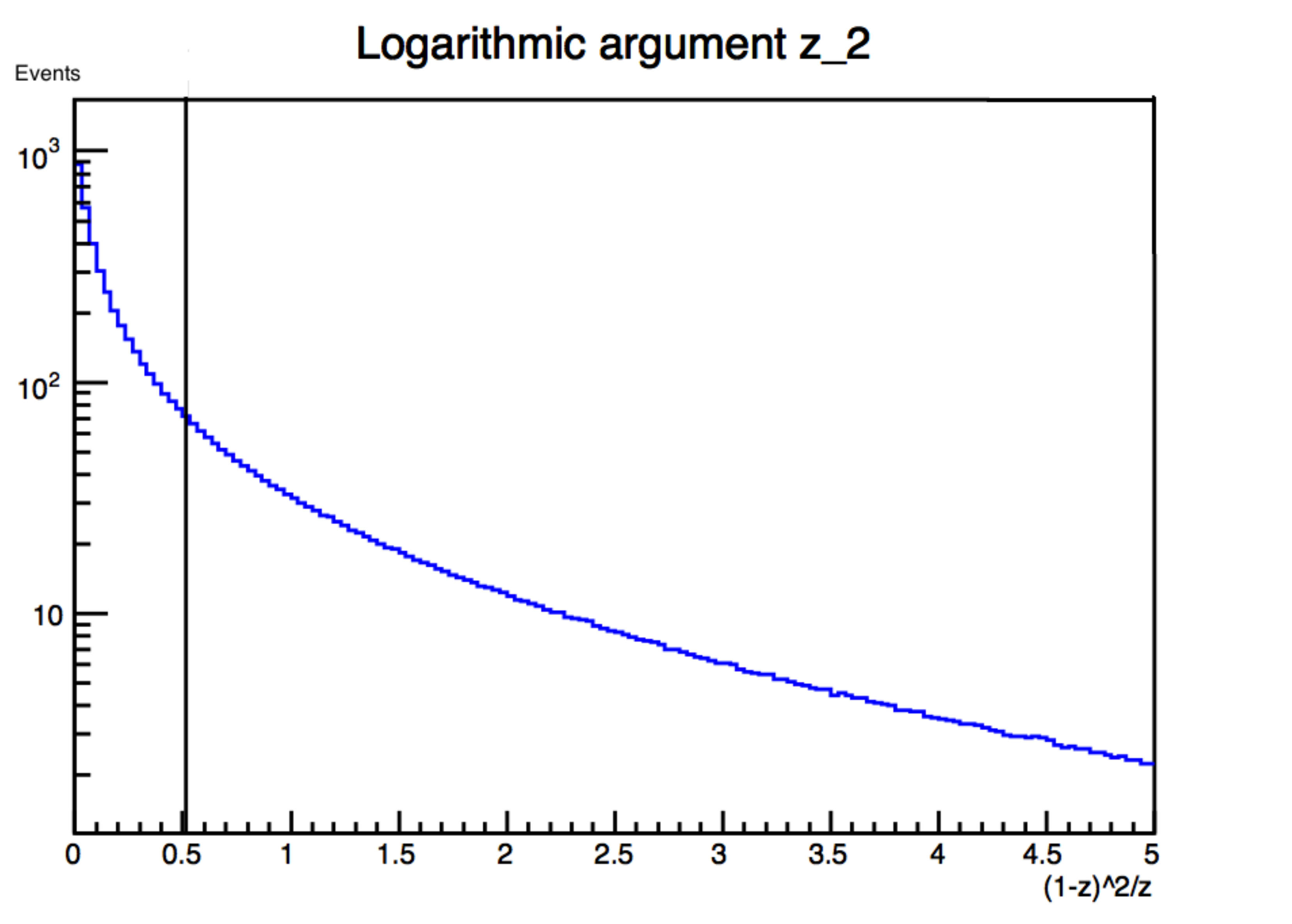}
\end{center}
\caption{Associate $H$ and bottom pair production at the LHC 7
  TeV. $m_H=126$. The distribution of the events (in pb/bin) as a
  function of ${\cal Q}^2_{\rm DY}(z)/m_H^2=(1-z_1)^2/z_1$ (left)
and ${\cal Q}^2_{\rm DY}(z)/m_H^2=(1-z_2)^2/z_2$ (right) are displayed.
 68\% of the events lie on the left of the vertical line.
\label{logs}}
 \end{figure}

The analysis presented in this contribution naturally leads to a number of
relevant follow-up studies. Up to now, we focussed on processes
featuring a single bottom in the initial state.
Processes that can be described by two $b$ quarks
in the initial state, such as $pp\to Hbb $ and $pp \to Z bb$
entail the first simple extension of the approach, which we are currently
working on~\cite{lmru}. The analytical expression of the leading order 
4FS cross section for both processes in the collinear limit 
is proportional to the product of two logarithms in the 
form of $L_{\rm DY}$, defined in terms of
two partonic variables $z_1$ and $z_2$ which depend on the partonic cross section,
the mass of the produced particle, $m_Z$ or $m_H$, and the invariant mass of 
the bottom pair. The distribution of the factors that multiply the
hard scales of the process in the logarithms are displayed in Fig.~\ref{logs}.
The suppression of scale of the logarithms with respect to the 
hard scale of the process due to the 
kinematical phase space factor is apparent. The findings are similar to the ones 
presented in this contribution and are generalised to the production of heavier particles
at future colliders.

To conclude, two major lines of research that we are pursuing are: the generalisation of
 the analysis from the level of inclusive cross sections to the level of differential distributions and
 the study of processes featuring heavy quarks in the final state. In the latter case fragmentation functions 
 play the role of parton distribution functions and the final state collinear logarithms are
 resummed by the time-like DGLAP evolution equations. 
 A thorough understanding of differential distributions, of the matching of fixed-oder
 calculations involving bottom quarks with parton showers and
 an assessment of the effect of the resummation of final state collinear logarithms would complete the picture 
 and provide a clear overview over a broad class of phenomenologically relevant processes.

\end{document}